\documentclass{phc-proc4-auth}
\usepackage{graphicx}
\usepackage{amssymb}

\def\a{\alpha}
\def\e{\varepsilon}
\def\d{\delta}
\def\t{\tau}
\def\n{\nu}
\def\o{\omega}
\def\s{\sigma}
\def\G{\Gamma}
\def\ra{\rightarrow}

\def\pd{\partial}

\def\om{\omega_m}

\def\ot{\tilde\omega}

\def\DD{{\mathcal D}}

\def\be{\begin{equation}}\def\ee{\end{equation}} 
\def\bea{\begin{eqnarray}}\def\eea{\end{eqnarray}} 
 
\def\lb{\label} 
\def\pref#1{(\ref{#1})}

\begin{document}

\begin{frontmatter}
\author{L.\ Benfatto and C.\ Morais Smith }
\address{D\'epartement de Physique, Universit\'e de Fribourg, 
P\'erolles, CH-1700 Fribourg, Switzerland}



\title{Optical response for a discrete stripe}


\begin{abstract}
The optical response of a charge stripe in the presence of pinning
impurities is investigated. We address the issue of a discrete  
description of the stripe, and discuss its quantitative relevance with
respect to a continuum one 
in the light of recent optical-conductivity measurements
in cuprate compounds.
\end{abstract}

\end{frontmatter}

% main text

The role of charge inhomogeneities in doped two-dimensional antiferromagnets
has been recently the subject of intense experimental and theoretical
investigations \cite{rew}. In particular, the physics of high-$T_c$ cuprates
stimulated the interest on theoretical models for charge segregation 
in one-dimensional stripes acting as domain walls for the surrounding
antiferromagnetism. In the present paper we study the optical
response of a string of holes (stripe) in the presence of pinning impurities. 

Let us consider a single vertical line of holes enumerated by $n$ on a square 
lattice with lattice constant $a$. The holes can move only in the
transversal ($x$) direction. The phenomenological Hamiltonian 
describing the system is
\be
H=\sum_n\left[ -t\cos ({p_n a})+\frac{J}{2 a^2}(u_{n+1}-u_n)^2\right]
\lb{eq1}
\ee
where $t$ is the hopping parameter, $a$ is the lattice constant, $u_n$ is
the displacement of the n-th hole from the equilibrium position, $p_n$ is
its conjugate transversal momentum, and $J$ is the stripe stiffness,
determined by the surrounding antiferromagnetism. We put $\hbar=k_B=1$. As
it has been shown in Ref. \cite{prl99}, at
$t/J>4/\pi^2$ the stripe depins from the lattice.
In this regime we can
expand the cos-term as $\cos(p_na)\sim const.+(p_na)^2/2$, so that the
corresponding action for the dimensionless field $\psi_n=u_n/a$ reads:
\be
S_0=\frac{1}{2at}\sum_n\int d\t \left[\left(\frac{\partial\psi}
{\partial\tau}\right)^2 +(Jt)(\psi_{n+1}-\psi_n)^2\right].
\lb{eq2}
\ee
The (bare) Green function for the $\psi$ field reads:
\be
\DD_0(q,\o_m)=(at)[\o_m^2+\o_J^2\sin^2(qa/2)]^{-1},
\lb{eq3}
\ee 
with $\o_J=2\sqrt{Jt}$. Thus, in the long-wavelength limit ($q\ra0$) the 
$\psi$ field displays a sound-like behavior $\o=vq$ with the velocity
$v=a\o_J/2$. When the electric
field is applied  perpendicular to the stripe, a current arises
$J=e\sum_n(\pd \psi_n/\pd t)$, and the transversal wandering
of the holes shows perfect Fr\"ohlich's conductivity:
\be
\Re \s(\o)=-e^2 \o \Im \DD_0(q=0,i\om\ra \o-i\e),
\lb{eq4} 
\ee
where the analytical continuation of the Green function in the lower
half-plane has been performed.  In Ref. \cite{noi}, by assuming from the
beginning a continuum ($a\ra 0$) description for the field $\psi_n\ra
\psi(y)$ and $\psi_{n+1}-\psi_n\ra \pd\psi/\pd y$, it was shown that the
interaction with the impurities destroys the perfect conductivity $\Re
\s(\o)=e^2\pi at\d(\o)$ of the clean case, and $\s(\o)$ displays a peak at
finite frequency.  The value of this 
frequency was evaluated by means of a diagrammatic approach for the
disorder, both in the weak- and strong-pinning regimes. Here we investigate
how these results are modified within a discrete description for the
stripe.

We represent the pinning by the impurities as a parabolic confining
potential, provided by the scattering centers located at coordinates $i$
along the stripes, with strength $V_0$, {\it i.e.} 
$
S=S_0+(V_0/2)\int d\t \sum_i(\psi_i-\beta_i)^2.
$
We describe the full, time-dependent solution
$\psi_n(t)=\phi_n(t)+\psi_n^0$ as a fluctuation around the equilibrium
configuration $\psi_n^0$ which the hole would assume in the absence of
dynamical fluctuations. As a consequence, Eq. \pref{eq3}-\pref{eq4} will
refer now to the dressed Green function $\DD$ of the fluctuating field
$\phi_n$, evaluated according to the Dyson equation as
$\DD^{-1}(q,\om)=\DD_0^{-1}(q,\om)-\Sigma=(\om^2+v^2 q^2-\Gamma t)/(at)$,
where we have performed a rescaling $\Sigma=\Gamma/a$ such that $\Gamma$
has dimensions of energy. In Ref. \cite{noi} it has been suggested that the
weak-pinning regime is the relevant one to describe stripe pinning in
cuprates. In this regime, $\G$ can be evaluated in the Born approximation
as 
$\Gamma(\om)=-(V_0/{2})n_ia+({V_0}/{2})^2n_i a L^{-1}
\sum_q D_0(q,\om),
$
where $\DD_0$ was used for calculating the second-order correction and the
density of impurities $n_i$ arises after averaging over the impurity
positions. On defining the frequency $\o_0=n_i v=n_ia\sqrt{Jt}$, the
dimensionless quantities  $\ot=\o/\o_0$ and $G=\Gamma t/\o_0^2$, 
the parameter $\a=V_0/(2n_i aJ)$, and performing the analytical 
continuation, the equation for $\G$ takes the form
\bea
\lb
{eq10}
G(\o)&=& -\a-i\frac{\a^2}{2\ot}\frac{\o_J}{\sqrt{\o_J^2-\o^2}}, 
\quad \o<\o_J\\ 
\lb{eq11}
G(\o)&=& -\a-\frac{\a^2}{2\ot}\frac{\o_J}{\sqrt{\o^2-\o_J^2}}, \quad \o>\o_J.
\eea
The interaction with impurities generates both a real and an imaginary part
in the self-energy $G=G'+iG''$. The real part $G'$ determines the energy 
at which the zero-energy delta-like peak of the optical conductivity
shifts, and $G''$ controls the spreading of the peak around this
value. Indeed, according to Eq. \pref{eq4}, the real part of the 
optical conductivity is
$\Re\s(\o)=-\s_0\ot G''/[(\ot^2+G')^2+G''^2],$
where $\s_0=e^2 at/{\o_0}$. From Eq.s\
\pref{eq10}-\pref{eq11} one
sees that at $\o>\o_J$ $G''$ vanishes and consequently also
$\s(\o)$. This result is a consequence of assuming a finite lower cut-off
$a$ for the lengths: this reflects in turn in an upper bound for the
momenta and consequently for the frequencies. On the other hand, the
relevant physical processes occur at an energy scale lower than $\o_J$, as
we will discuss below. In particular, when $\o\ll\o_J$ Eq. \pref{eq10}
reduces to
\be
G= -\a-\frac{i\a^2}{2\ot}.
\lb{eq12}
\ee
which is exactly the result obtained in Ref. \cite{noi}. In such a case,
the resulting conductivity reads
\be
\Re\s(\o)=\s_0\frac{2(\o/\n)^2}{4(\o/\n)^2[(\o/\n)^2-1]^2+\a},
\lb{eq13}
\ee
with $\n=\sqrt{\a}\o_0$. We note that $\Re\s(\o)\ra 0$ both as $\o\ra 0$
and as $\o\ra \infty$, and has a peak at $\o\simeq \n$. The optical
conductivity $\s(\o)$ evaluated with the self-energy \pref{eq10}-\pref{eq11}
shows remarkable differences from Eq.\ \pref{eq13} only when $\nu\sim\o_J$,
but as we shall discuss below this case is not physically relevant because
in any case $\nu<\o_J$. Dimensional estimates of Eq. \pref{eq2} indicate
that  the effect of pinning can be
understood in terms of trapping the sound mode on a finite length scale
$\lambda$, so that $\n=v/\lambda$.  In the weak-pinning case ($\a\ll1$)
considered here $\lambda$ is the Larkin-Ovchinnikov length. Because 
$\lambda$ cannot be smaller than the lattice spacing $a$, $\nu<\o_J$
always. 

Let us discuss now the physical values of the parameters in comparison with
recent measurements of the infra-red response of La-based cuprates
\cite{cu1}. The optical conductivity of these compounds displays a huge
peak in the far infra-red region at a frequency around $5-25$ meV
(depending on doping), which we attributed to the pinning of transversal
stripe wandering \cite{noi}. For cuprates, physical values of the parameters
are $t\sim J\sim 0.1$ eV. By considering the dopants themselves as
scattering centers, we can also estimate on average $n_ia=0.5$
\cite{noi}. This means that $\o_0\sim 50$ meV, and because $n_ia\sim 1$
corresponds to weak pinning, we expect that the peak frequency
$\nu=\sqrt{\a}\o_0$ is further reduced with respect to this value, in
agreement with experimental data.  Note that for these physical values of
$t,J$: (i) $t/J\sim 1$ so that the stripe is depinned from the lattice;
(ii) $\o_J=200$ meV, thus $\nu$ is always found
in the range where $\nu\ll\o_J$ and as a consequence Eq. \pref{eq12} is a
good approximation for the self-energy \pref{eq10}.

In summary, we evaluated the optical response of a discrete hole stripe in
the presence impurities. The impurity centers pin the sound-like mode
associated to the transversal motion of the holes. By retaining the lattice
parameter $a$ as a lower cut-off for the length we find an upper limit for
the optical response of the system. However, for parameter values
relevant for cuprates the pinning process always takes place at energies
$\nu\ll\o_J$, where the details of the discrete lattice description can be
neglected and a continuum ($a\ra 0$) description for the elastic stripe can
be adopted.

This work was supported by the Swiss National Foundation for Scientific 
Research under grant No.~620-62868.00.

\end{document}